\documentclass{webofc}
\usepackage[varg]{txfonts}   
\usepackage{xcolor}
\usepackage{lineno}

\begin{document}
\title{ALICE Central Trigger System for LHC Run 3}
%
%

\author{\firstname{Jakub} \lastname{Kvapil}\inst{1}\fnsep\thanks{\email{jakub.kvapil@cern.ch}} \and
    	\firstname{Anju} \lastname{Bhasin}\inst{2,4}\and
    	\firstname{Marek} \lastname{Bombara}\inst{3}\and
        \firstname{David} \lastname{Evans}\inst{1}\and
        \firstname{Anton} \lastname{Jusko}\inst{1}\and
        \firstname{Alexander} \lastname{Kluge}\inst{4}\and
        \firstname{Marian} \lastname{Krivda}\inst{1}\and
        \firstname{Ivan} \lastname{Kralik}\inst{5}\and
        \firstname{Roman} \lastname{Lietava}\inst{1}\and
        \firstname{Sanket Kumar} \lastname{Nayak}\inst{1}\and
        \firstname{Simone} \lastname{Ragoni}\inst{1}\and
        \firstname{Orlando} \lastname{Villalobos Baillie}\inst{1}
}

\institute{University of Birmingham, Birmingham, UK
\and
Physics Department, University of Jammu, India
\and
           Pavol Jozef Safarik University, Kosice, Slovakia
           \and
           European Organization for Nuclear Research (CERN), Switzerland
\and
           Slovak Academy of Science, Kosice, Slovakia
          }

\abstract{%
  A major upgrade of the ALICE experiment is in progress and will result in high-rate data taking during LHC Run 3 (2022-2024).
  The LHC interaction rate at Point 2 where the ALICE experiment is located will be increased to $50\ \mathrm{kHz}$ in Pb--Pb collisions and $1\ \mathrm{MHz}$ in pp collisions. The ALICE experiment will be able to read out data at these interaction rates leading to an increase of the collected luminosity by a factor of up to about 100 with respect to  LHC Runs 1 and 2. To satisfy these requirements, a new readout system has been developed for most of the ALICE detectors, allowing the full readout of the data at the required interaction rates without the need for a hardware trigger selection. A novel trigger and timing distribution system will be implemented, based on Passive Optical Network (PON) and GigaBit Transceiver (GBT) technology. To assure backward compatibility a triggered mode based on RD12 Trigger-Timing-Control (TTC) technology, as  used in the previous LHC runs, will be maintained and re-implemented under the new Central Trigger System (CTS). A new universal ALICE Trigger Board (ATB) based on the Xilinx Kintex Ultrascale FPGA has been designed to function as a Central Trigger Processor (CTP), Local Trigger Unit (LTU), and monitoring interfaces.

  In this paper, this new hybrid multilevel system with continuous readout will be described, together with the triggering mechanism and algorithms. An overview of the CTS, the design of the ATB and the different communication protocols will be presented.

}
\maketitle
\section{Introduction}\label{intro}

A Large Ion Collider Experiment (ALICE) \cite{alice} at the Large Hadron Collider (LHC)~\cite{LHC} started its operation in 2008 and has produced important results in both heavy-ion and proton-proton collisions. The ALICE Central Trigger Processor (CTP) \cite{CTPTDR} used in LHC Run 2 operated with four trigger latencies LM ($650\ \mathrm{ns}$), L0 ($900\ \mathrm{ns}$), L1 ($6.5\ \mathrm{\mu s}$), and L2 ($88\ \mathrm{\mu s}$). The LM level was sent only to the Transition Radiation Detector (TRD), 
which therefore received all four levels, while all other detectors received L0 as their first trigger signal. The LM board design was used as an alpha version of the future ALICE Trigger Board (ATB). The CTP consisted of 6 different 6U Versa Module Europa (VME) boards, with in addition a Local Trigger Unit (LTU) board for each detector. Communication between boards was done via a customised back-plane. The communication between LTUs and detector front-end electronics (FEE) was done using the Trigger-Timing-Control (TTC) protocol developed by the RD12 collaboration \cite{RD12}. The designed interaction rate was $8\ \mathrm{kHz}$ for Pb--Pb and $100\ \mathrm{kHz}$ for pp collisions. The maximum readout rate was around 1 kHz due to the space-charge and bandwidth limitations of the Time Projection Chamber detector.

For LHC Run 3, ALICE will upgrade the readout system to implement continuous detector readout. The ALICE detectors will be self-triggered, constantly pushing the data stream, and the CTP will provide time-stamps to synchronise data from different detectors. The original idea of the triggered event will no longer be valid, as there are no discrete events but rather a continuous stream. The interaction rate will be $50\ \mathrm{kHz}$ for Pb--Pb, $500\ \mathrm{kHz}$ for p--Pb, and $1\ \mathrm{MHz}$ for pp collisions and every event will be read out. To incorporate these changes a new CTP and LTUs were developed for timing distribution alongside the new Common Readout Unit (CRU) \cite{CRU} for the readout. However, since some detectors are not upgrading their hardware, a trigger distribution and the RD12 TTC protocol must also be available.


\section{Overview of Central Trigger System}\label{CTS}
The layout of the new Central Trigger System (CTS) is similar to the old one. One CTP board  distributes the  clock and triggers to 18 LTU boards, one for each ALICE detector, one in test setup and two spare boards. During normal (global) data-taking operations the CTP  controls the LTUs, which act as a communication interface with the detector CRUs. Due to the two-tier system, a single LTU can also be decoupled from the CTP and the other LTUs and operated in the so-called standalone mode to perform independent tests without influencing the data-taking of the other detectors. In standalone mode the LTU can emulate the presence of the CTP. This feature is particularly important for the commissioning phase.

\begin{figure}[ht]
	\centering
	\includegraphics[width=0.75\textwidth,clip]{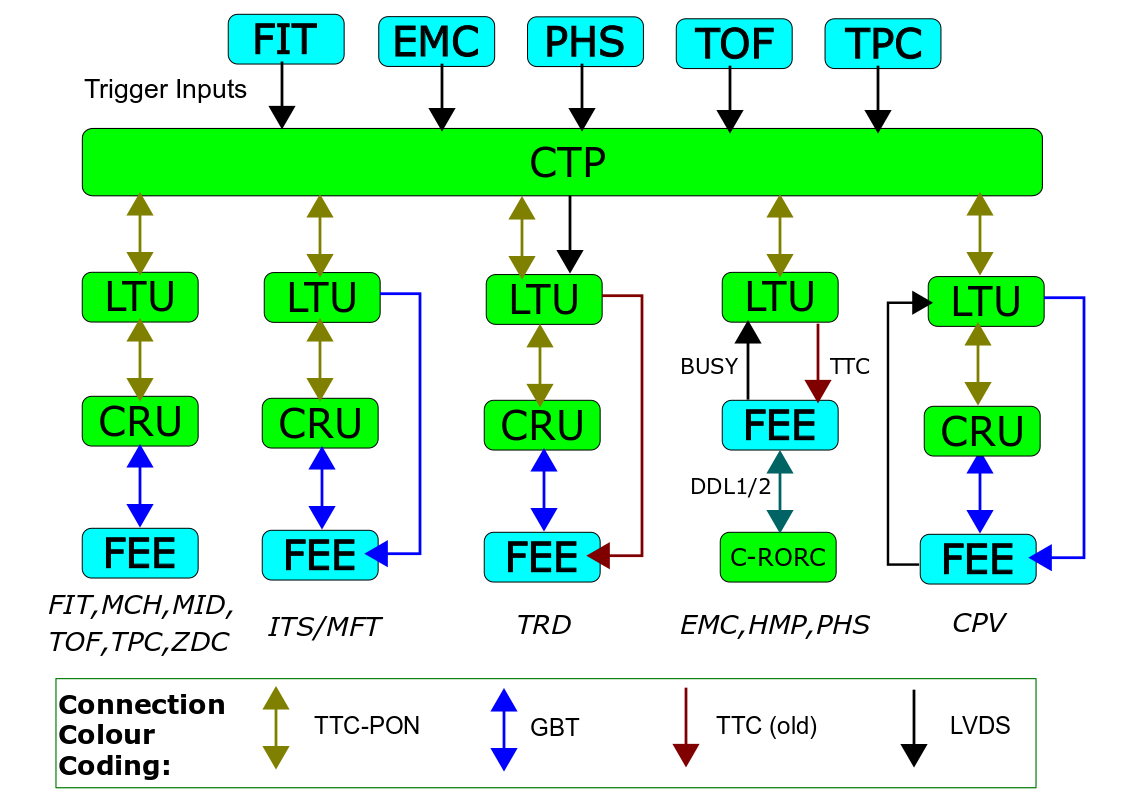}
	\caption{Central Trigger System overview showing trigger inputs to the CTP and the connection types between the CTP and the detector FEE.}
	\label{CTSdist}       
\end{figure}

An overview of the ALICE CTS is shown in Figure~\ref{CTSdist}. There are several hardware protocols used for CTS communication with detectors. Detectors that are fully upgrading their FEE, namely the Fast Interaction Trigger (FIT), which is composed of the 3 sub-detectors FT0, FV0 and FDD, Muon Chambers (MCH), Muon Identifier (MID), Time-of-Flight (TOF), Time Projection Chamber (TPC), Inner Tracking System (ITS), Muon Forward Tracker (MFT) and Zero Degree Calorimeter (ZDC), all use the Common Readout Unit card for readout. The link between LTU and CRU  uses the bidirectional TTC Passive Optical Network (PON) \cite{PON} technology and the communication between CRU and detector FEE  uses the GigaBit Transceiver (GBT) \cite{GBT} technology. In addition to these connections, the ITS and MFT require a direct GBT link between LTU and FEE. The Electromagnetic Calorimeter (EMC), Photon Spectrometer (PHS), and High Momentum Particle Identification (HMP) are not upgrading their electronics and will continue using the RD12 TTC protocol for downstream communication. The Transition Radiation Detector (TRD) uses the TTC-PON system but requires triggers over RD12 TTC optical lines. The Charged Particle Veto (CPV) receives clock and trigger signals over GBT links. The EMC, PHS, HMP, TRD and CPV cannot read out data at the required rate and each of them will generate a BUSY signal to inform the CTP when additional triggers cannot be received.

\section{The Time and Trigger distribution in ALICE}\label{contin}
Starting from the LHC Run 3, the ALICE default operation mode will be in continuous readout at a rate of $50$ kHz in Pb--Pb collisions and $1$ MHz in pp collisions. A subsequent data selection is done using software filters. The detectors using CRUs must also be able to operate in triggered mode, which is especially important during the commissioning phase. Detectors not using CRU and GBT links are operated in triggered mode, with the CTP providing triggers distributed via the RD12 TTC protocol as explained above. Detectors using the RD12 TTC protocol and the CPV are triggered using a two-level scheme with the possibility to operate using trigger signals on three different latencies (LM, L0, L1). The two-level system is particularly important for detectors using the RD12 TTC protocol due to the limited readout bandwidth.

The trigger input detectors listed in Table \ref{tab:trgin}  provide a total of 39 trigger inputs to the CTP on different latencies. In addition, the LHC Beam Pick-up Timing system (BPTX) will contribute with 2 inputs on L0 latency. Some detectors can provide multiple inputs corresponding to different trigger conditions. These inputs are only important for detectors requiring triggers.

\begin{table}[ht]
	\centering
	\caption{CTP trigger inputs with their latency and contributing trigger level.}
	\label{tab:trgin}       
	\begin{tabular}{|c|c|c|c|}
		\hline
		Detector & Lat./Level & N of inputs & Latency [ns]  \\\hline
		FT0 & LM & 5 & 425 \\
		FV0 & LM & 5 & 425 \\
		FDD & L0 & 5 & - \\
		TOF & L0 & 4 & 862 \\
		EMC & L0 & 2 & 843 \\
		& L1 & 8 & 6100 \\
		PHS & L0 & 2 & 843 \\
		& L1 & 5 & 6100 \\
		BPTX & L0 & 2 & - \\
		TPC & L0 & 1 & 900 \\\hline
	\end{tabular}
\end{table}

The ALICE data are divided into so-called HeartBeat frames (HBf) and 128 (programmable value) of them will compose a Time Frame (TF). A HBf is set to have the same length as an LHC orbit ($\approx 88.92\ \mathrm{\mu s}$), which is approximately the time required to fully read out the hits from the TPC detector. The CTP sends a HeartBeat (HB) signal and a HeartBeat reject (HBr) flag at the beginning of each HBf. The CRU either accepts or rejects data in the upcoming HBf according to the HBr flag. In response, the CRU  sends back a HeartBeat acknowledge message (HBam), containing a HeartBeat acknowledge (HBack) flag for whenever an HBf was successfully transferred to the First Level Processor (FLP) \cite{CRU}. Since the HBfs are buffered, successful transmission can happen up to 8 HBf after it was collected. The HBam also carries the status of the CRU buffers (Buffers Status (BS)). The HBr allows control over CRU throttling and is used in three ways. In the first case, the CTP does not issue any HBr and every CRU runs in autonomous mode. In the second case, the HBr is used to downscale the data rate by generating a certain number of HBr signals in a TF. Lastly, the CTP evaluates the CRU buffer statuses every HBf and decides whether to generate a HBr or not - this is called the collective mode.

\begin{figure}[ht]
	\centering
	\includegraphics[width=0.7\textwidth,clip]{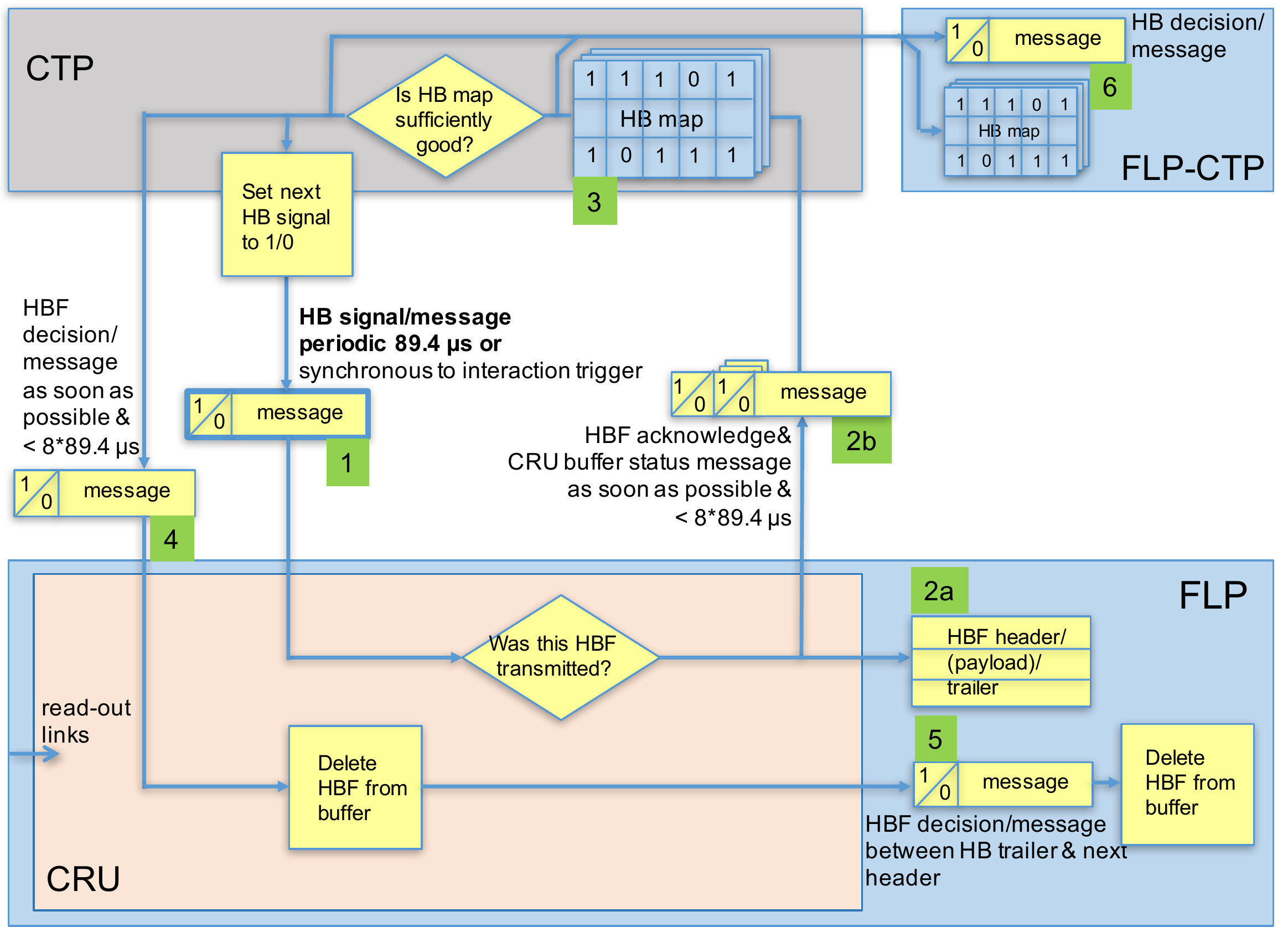}
	\caption{A logical flow of HeartBeat signals between CTP, CRU and FLP.}
	\label{HBflow}       
\end{figure}

The CTP collects the HBack flags from all 441 CRUs and builds a so-called Global HeartBeat Map (GHBmap). The HeartBeat decision (HBd) is calculated using a HB function $f$. Since it is not possible to implement a general function of $N=441$ bits, the problem is simplified in the following way. A HeartBeat Mask (HBmask) corresponding to a certain subset of CRUs is introduced and the function $f$ is defined as
\begin{equation}\label{key}
HBd = f(GHBmap \wedge HBmask),
\end{equation}
where $\wedge$ represents a logical "and" and the function $f$ is a simple logical function of the "and" result. The HBd is then asynchronously transmitted via CRUs to FLP to notify them whether to save or discard a particular HBf. There will be a HB function for each detector to be universal in the creation of detector groups (clusters). For example, if the function is the same for each detector, ALICE behaves as one detector for readout, whereas if a different function is used for each detector, then each detector works independently of the others. A schematic diagram of the HB logic flow can be seen in Figure \ref{HBflow}.

For the detectors that can run only in triggered mode, a design of trigger classes, similar to Run 2, is used. A trigger class is defined as a set consisting  of trigger conditions, cluster, and vetoes. A trigger condition contains a logical function of trigger inputs and a Bunch Crossing (BC) mask (which specifies which BC slots correspond to  physical bunch crossings). A trigger cluster is a group of detectors to be read out together. The trigger veto consists of the logical "or" of a cluster's detector busy and down-scaling veto. There can be up to 64 classes and 18 clusters in Run 3. Since triggered detectors usually need to be read out together with detectors in continuous mode, a HBr of those detectors will act as another busy veto, so the CTP does not send triggers during HBr windows. Moreover, triggered detectors do not contribute to the HBd evaluation as the HBd is evaluated long after the triggered detectors store data. This design allows independent operation of continuous readout detectors and a partial correlation with triggered detectors.

\section{Implementation of the CTS}

\subsection{Trigger Protocols and Transceivers}\label{protocol}
The CTP generates a timing and trigger message which is transmitted to the FEE to synchronise all ALICE detectors. The trigger message  consists of 80 bits arranged as follows: trigger type (32 bits), BC counter (12 bits), trigger level (4 bits), Orbit counter (32 bits). The bunch crossing number and orbit counters make up the event identification used to time stamp the data. The trigger time carries information about start/end of run, HeartBeat, physics, orbits, calibration etc.

The Passive Optical Network (PON) is an off-the-shelf technology using the Optical Line Terminal (OLT) acting mainly as a transmitter, and the Optical Network Unit (ONU), acting mainly as a receiver. It allows up to 9.6 Gbps data flow. Both units can transmit and receive, however, the opposite transceiver has limited bandwidth and speed. Each network link can send 30 8-bit words per bunch crossing ($24.95\ \mathrm{ns}$), of which five are for PON internal use, leaving 200 user-defined bits. The first 15 words are used for trigger messages together with additional monitoring. The last 10 bytes are used to transmit the HeartBeat Decision Record (HBDR) of a particular TF as described in Section \ref{contin}.

The GBT protocol allows bidirectional transmission directly to the detector FEE. An LTU sends 120 bits per bunch crossing; the first 80 bits are the trigger message, identical to the PON message. The remaining bits are reserved for GBT internal use.


The RD12 TTC consists of 2 channels allowing transmission of only one bit per bunch crossing per channel. The TTC-A channel is used to transmit a synchronous trigger for two levels (1-bit for LL0 and 2-bit for LL1) and TTC-B is for asynchronous trigger messages. Each channel is Manchester-encoded with the clock and they are multiplexed into a $160.316\ \mathrm{MHz}$ optical line by the TTCex and sent to the detector FEE. One TTC-B word is a 42 bits long Hamming-encoded message, where 16 bits are available for the user, of which 4 bits are reserved for the word header, effectively leaving 12 data bits. The 76 bits of trigger message (excluding the trigger level which is sent over TTC-A) is split between 7 TTC-B words limiting the maximum trigger rate to about $130\ \mathrm{kHz}$, sufficient for TTC detectors. In addition, each LHC orbit and any calibration requests, both used for synchronisation, are sent synchronously in TTC-B as a short broadcast (16 bits - 6 for user) command. To allow synchronous transmission in a channel that is  by default asynchronous, a re-synchronisation is done for the duration of the broadcast transmission. Since the busy is treated on the CTP/LTU side, an internal busy is raised to prevent new trigger generation for the duration of L0-L1 to address detector busy propagation time towards the CTP/LTU. Lastly, since the transmitter bandwidth is limited, a derandomisation is implemented, and once the buffers are full, a busy is raised to prevent acceptance of new triggers.

\subsection{ALICE Trigger Board}\label{HW}
\begin{figure}[ht]
	\centering
	\sidecaption
	\includegraphics[width=0.69\textwidth,clip]{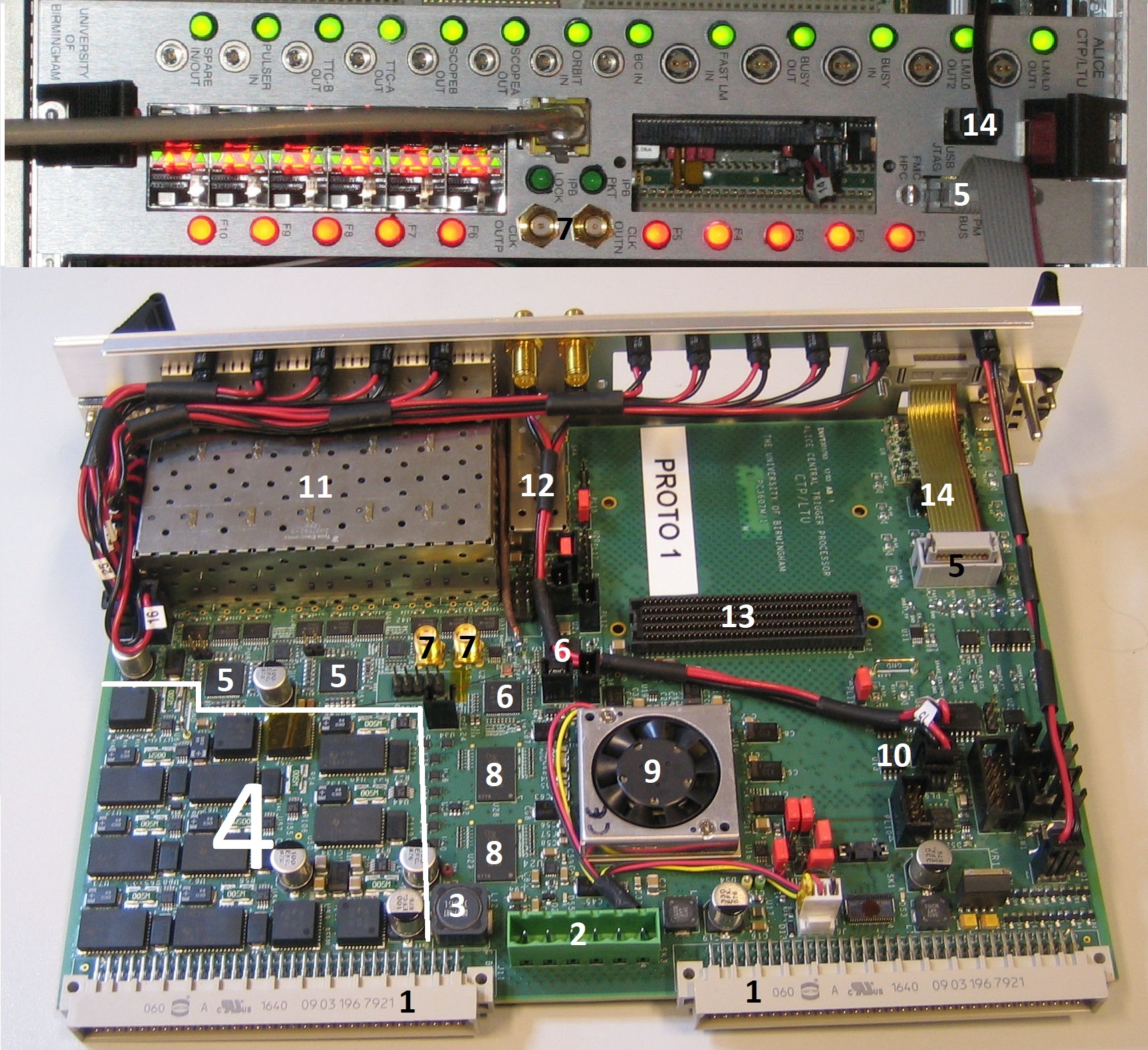}
	\caption{Front and side view of ALICE Trigger Board (ATB). The following components are visible (see in the text for details).\\
		1: VME 6U power supply connector\\
		2: ELMAbox power supply connector\\
		3: power decoupling - inductance\\
		4: DC-DC converter\\
		5: PMbus\\
		6: PLL\\
		7: Clock from PLL\\
		8: DDR4\\
		9: FPGA\\
		10: Flash memory\\
		11: two six-fold SFP+\\
		12: single-fold SFP+\\
		13: FMC\\
		14: JTAG-microUSB\\		
	}
	\label{board}       
\end{figure}

A general ALICE Trigger Board (ATB) was designed to act as CTP, LTU, LTU-GBTit (GBT interface test) and LTU-TTCit (TTC interface test) with small modifications. The board can be seen in Figure \ref{board}. The board can be powered either through the 6U VME crate using $+5$V/$10$A $\pm12$V/$1$A or using a standalone crate which provides the required voltages and cooling. The standalone box version has proved to be particularly useful when distributing LTU boards to detector groups for standalone test runs. From the power supply, the DC-DC converter generates 12 different power domains required for the board operation. To monitor the domains two UCD90120A Power Management Bus (PMBus) chips are present on the board with a connector on the front panel for monitoring using the $\mathrm{I^2C}$ protocol. Two Si5345 Phase-Locked Loops (PLL) are used to generate a very precise clock with a random jitter of 1 ps which can be locked to the LHC main clock using the TTC machine interface (TTCmi). The most strict requirement is from the ALICE TOF detector, requiring 12 ps jitter at worst. The main three clock domains are $240.474\ \mathrm{MHz}$, which serves as the central board clock, $40.079\ \mathrm{MHz}$ as the LHC bunch-crossing frequency, and $31.25\ \mathrm{MHz}$ for ethernet communication. Two 1 GB MT40A512M16HA-083E Double Data Rate 4 Synchronous Dynamic Random Access Memory (DDR4 SDRAM) are used to store signals used for monitoring and debugging. One DDR4 is used to monitor the inputs and the other for the outputs of the board. The LTU board is equipped with a Xilinx Kintex Ultrascale FPGA (XCKU040-2FFVA1156E) and flash memory (MT25QU128ABA1EW7) to store the FPGA logic. The CTP board is equipped with a more powerful FPGA (XCKU060-2FFVA1156E) and a bigger flash memory (MT25QU256ABA1EW7) to accommodate the larger CTP requirements  for logic. The flash memory can be programmed using a microUSB-JTAG connection from the front panel. A two-fold Small form-factor pluggable transceiver (SFP+) cage is used to plug in high-speed link optical modules and a single-fold SFP+ case for a plug-in ethernet module. The FPGA Mezzanine Card (FMC) port can be utilised to change the purpose of the general trigger board. The CTP FMC card will be used on the CTP board to utilise up to 64 LVDS trigger inputs/outputs. The FMC FM-S18 rev. E can be used as another SFP+ two-fold cage to increase the number of available optical links on LTU boards. Furthermore, by using FMC GBTx, the board can be converted into a LTU-GBTit board to monitor the machine interface clock and data on GBT links. Finally, by using the TTCrx FMC card the board will be converted into the LTU-TTCit board to monitor data on TTC links. On the front panel, there are 8 ECL LEMO 00B connectors for RD12 TTC, scope, clock, and orbit signals, 6 LVDS LEMO B connectors for triggers and busy and an SMA connector which is directly connected to PLL and provides the clock.

The general LTU board has one ONU to receive data from the CTP, nine OLT for connecting detector, one downstream GBT for CPV (so general firmware can be used) and 1 bidirectional GBT, in case each LTU is connected with the FLP via CRU for monitoring. The ITS and MFT detectors require more GBT links thus their board will be equipped with one ONU, one OLT, one bidirectional GBT, and 13 downstream GBT (of which 4 are located on  the FMC).
A summary of the detector operation latencies/trigger levels and optical and electrical connections between LTU, CRU and FEE is given in Table \ref{tab:detlata}.

The CTP board is equipped with 9 OLT to control the 18 LTU, and 3 bidirectional GBT to send the interaction record, trigger class record, and monitoring to the CRU-FLP. Using the GBT protocol for the connection to CRU-FLP, an LTU and CTP  behave as an additional detector for the readout system.

\begin{table}[ht]
	\centering
	\caption{A summary of operating levels, number of OLT links with optical attenuation and splitter, number of GBT and TTC links, LVDS with the level it operates on, and how the busy is propagated back to CTP for various ALICE detectors.}
	\label{tab:detlata}       
	\begin{tabular}{|c|c|c|c|c|c|c|c|c|}
		\hline
		Detectors & Lat./Level & OLT & opt. att & opt. split. & GBT & TTC & LVDS & BUSY   \\\hline
		MID& LM & 1 & 10 dB & 1:4 & - & - & - & -   \\
		TOF& LM& 1 & 10 dB & 1:4 & - & - & - & -   \\
		FT0& LM& 1 & 15 dB & none & - & - & - & -   \\
		FV0& LM& 1 & 15 dB & none & - & - & - & -   \\
		FDD& LM& 1 & 15 dB & none & - & - & - & -   \\
		ZDC& LM& 1 & 15 dB & none & - & - & - & -   \\
		MCH& LM& 1 & none & 1:32 & - & - & - & -   \\
		TPC& LM& 6 & none & 6x 1:64 & - & - & - & -   \\
		ITS& LM& 1 & none & 1:32 & 6/12 & - & - & -   \\
		MFT& LM& 1 & none & 1:32 & 6 & - & - & -   \\
		TRD& LM - L0& 9 & 10 dB & 9x 1:4 & - & 20 & - & TTC-PON   \\
		CPV& L0 - L1& 1 & 15 dB & none & 1 & - & LO & LVDS   \\
		HMP& LM - L1& - & - & - & - & 2 & LM & LVDS   \\
		PHS& L0 - L1& - & - & - & - & 2 & - & LVDS   \\
		EMC& L0 - L1& - & - & - & - & 3 & - & LVDS   \\\hline
		
	\end{tabular}
\end{table}

\begin{figure}[ht],
	\centering
	\begin{minipage}[b]{\textwidth}
		\includegraphics[width=\textwidth,clip]{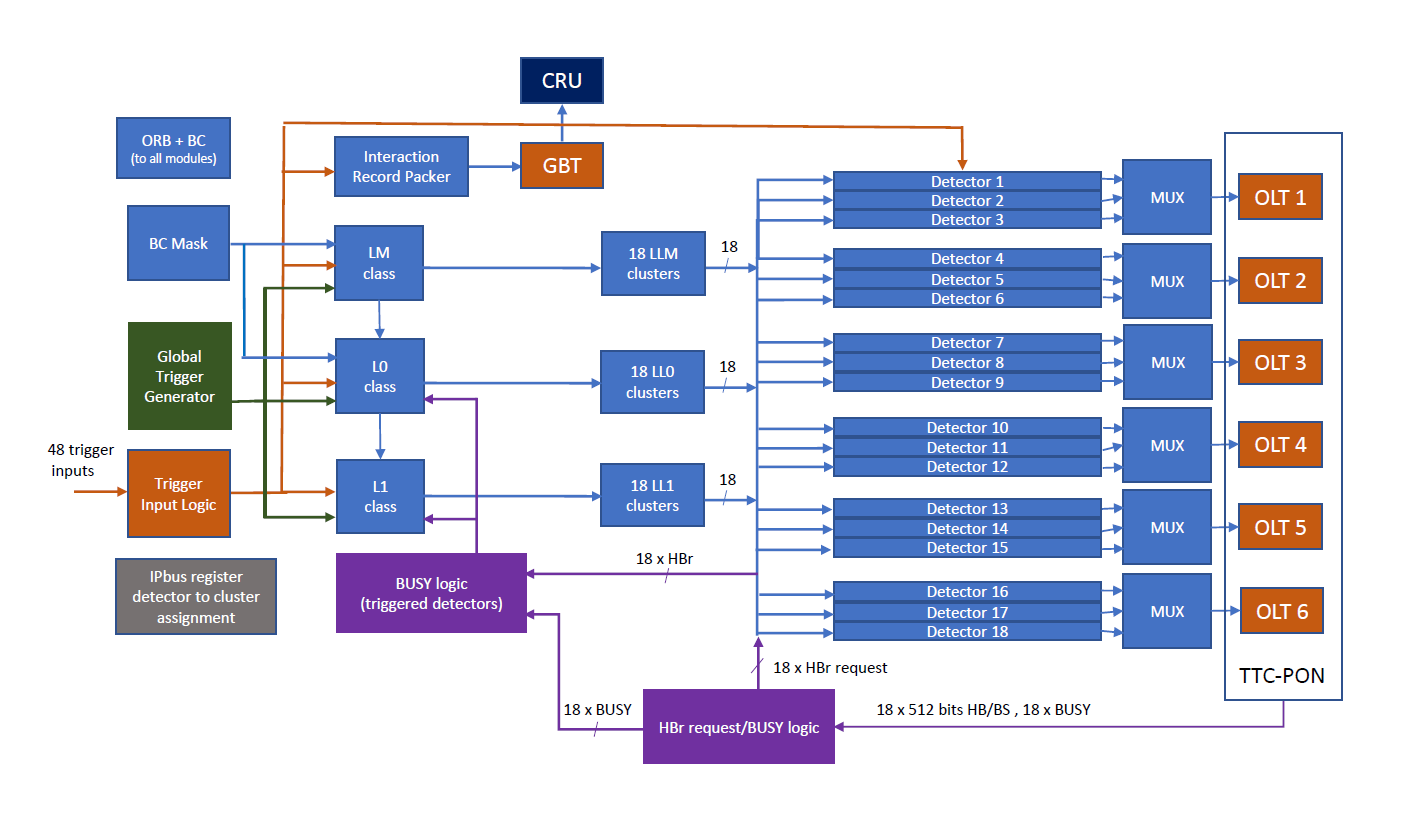}
	\end{minipage}
	\caption{Main elements on the CTP FPGA firmware.}
	\label{CTPLTUFW}       
\end{figure}

Both CTP and LTU firmware contain some common elements, namely ONU, OLT, GBT transceivers, IPbus \cite{IPbus} and drivers for DDR4 memories and peripherals. The IPbus is used for control and monitoring using Reverse Address Resolution Protocol (RARP) to get an IP address after power up. The CTP board can receive trigger inputs either from detectors or external emulator Trigger Input/Clock Generator (TICG), moreover, it can also emulate triggers from stored patterns in DDR4 memory functioning as Trigger Data Generator (TDG). A so-called Interaction Record (IR) containing Trigger Inputs is sent over GBT towards CRU for bookkeeping. The LTU board contains a CTP emulator that can be used to substitute CTP during standalone runs. A preliminary block diagram of CTP FPGA can be seen in Figure \ref{CTPLTUFW}.

\begin{figure}[ht]
	\centering
	\includegraphics[width=0.9\textwidth,clip]{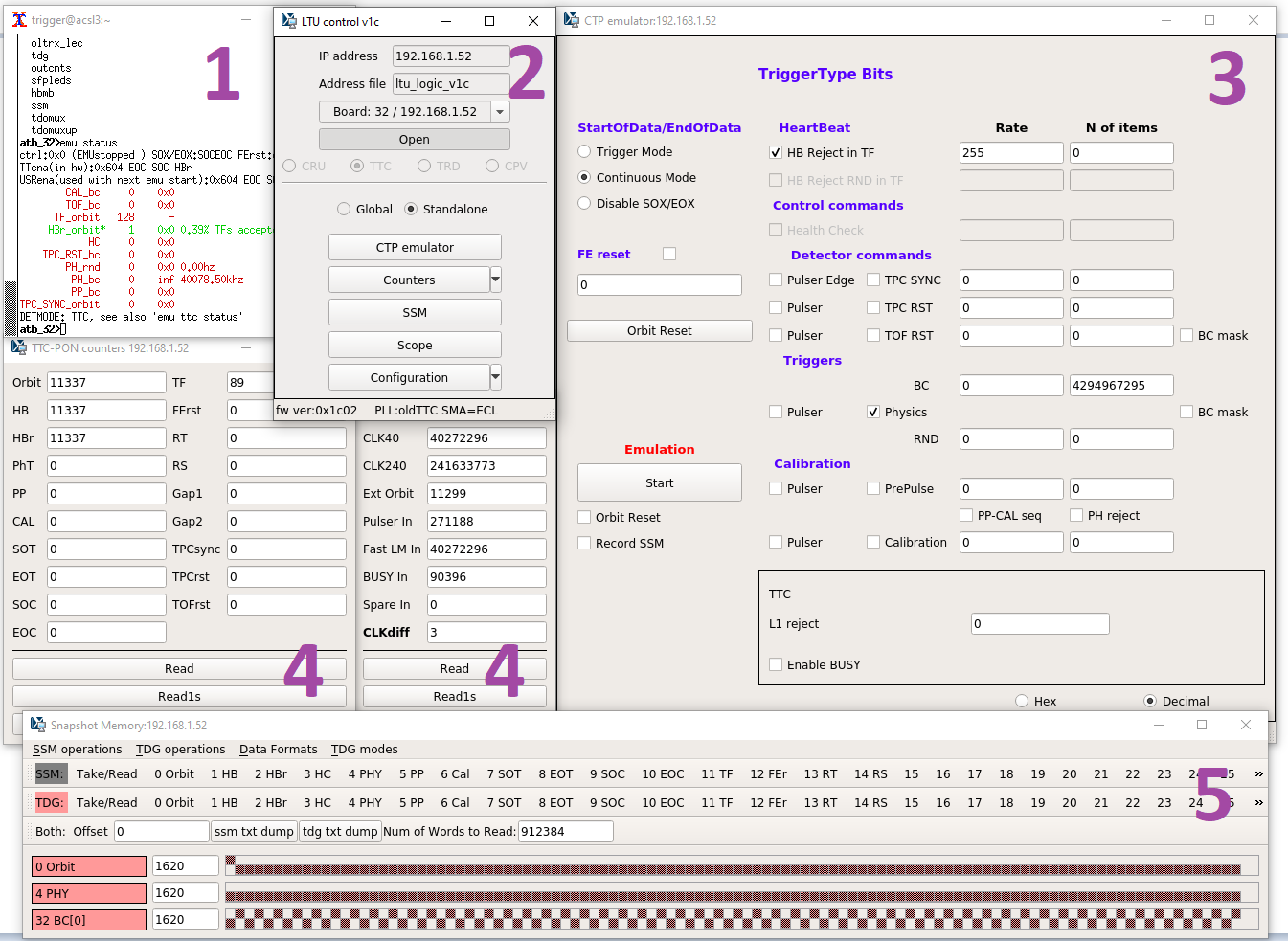}
	\caption{Overview of the possible control panels for both versions of the control and configuration software: Terminal provided from the Python-based system (1) Qt-based Graphical Interface with panels for configuration (2), trigger settings (3), counters (4) and DDR4 memory dump (5).}
	\label{QT}       
\end{figure}

For the control and configuration software, two independent systems have been developed, one for trigger experts and the other for detector experts. The ATB board is controlled in both systems through the IPbus suite \cite{IPbus}. The first system, written in Python, is used as a development and debugging tool for new firmware releases and user control, e.g. arranging particular triggers in board's CTP emulator. This tool is mainly used by trigger experts and only terminal control is supported and can be seen in Figure \ref{QT} (1). The second system offers selected parts of the control and a Graphical Interface based on C++ and Qt. This interface is mainly used by detector experts to configure and set trigger generation on the LTU and can be seen in Figure \ref{QT} (2-5). Lastly, during the global runs the CTP will be configured over the Experiment Control System (ECS) using the google Remote Procedure Calls (gRPC) protocol.

Four monitoring systems have been developed. A Detector Control System (DCS) project based on WinCC software \cite{WinCC} monitors the boards' and crates' temperature, voltages and currents using PMbus. A Quality Control (QC) system is used to monitor interaction and trigger rates and CTP data quality. To make sure the TTC-PON connection is working correctly, a Link Health Monitor (LHMon) has been implemented, in which the CRU sends a link counter that increases by 1 in each upstream transmission, and the receiving side verifies consistent increments. Lastly, a CTP-LTU monitoring system verifies the consistency of the CTP and LTU boards.

\subsection{Installation at LHC Point 2}
\begin{figure}[ht]
	\centering
	\begin{minipage}[b]{0.48\textwidth}
		\includegraphics[width=\textwidth,clip]{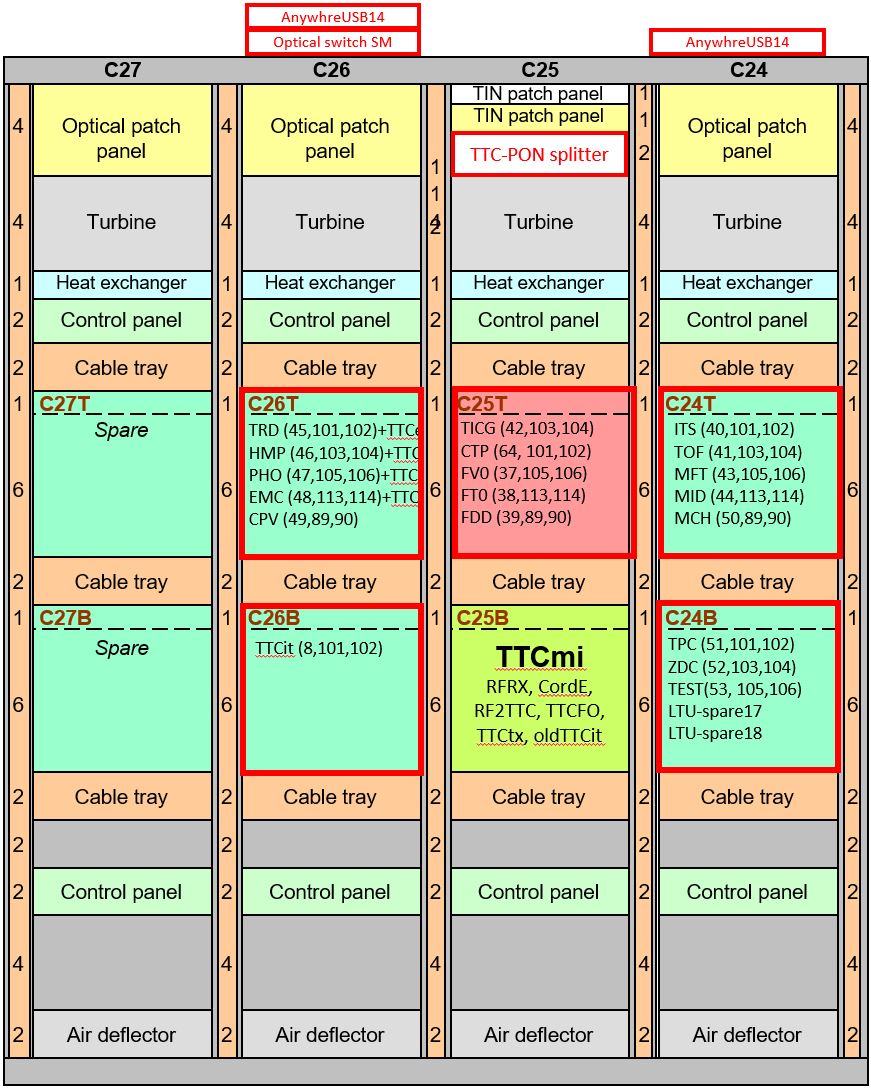}
	\end{minipage}
	\hfill
	\begin{minipage}[b]{0.45\textwidth}
		\includegraphics[width=\textwidth,clip]{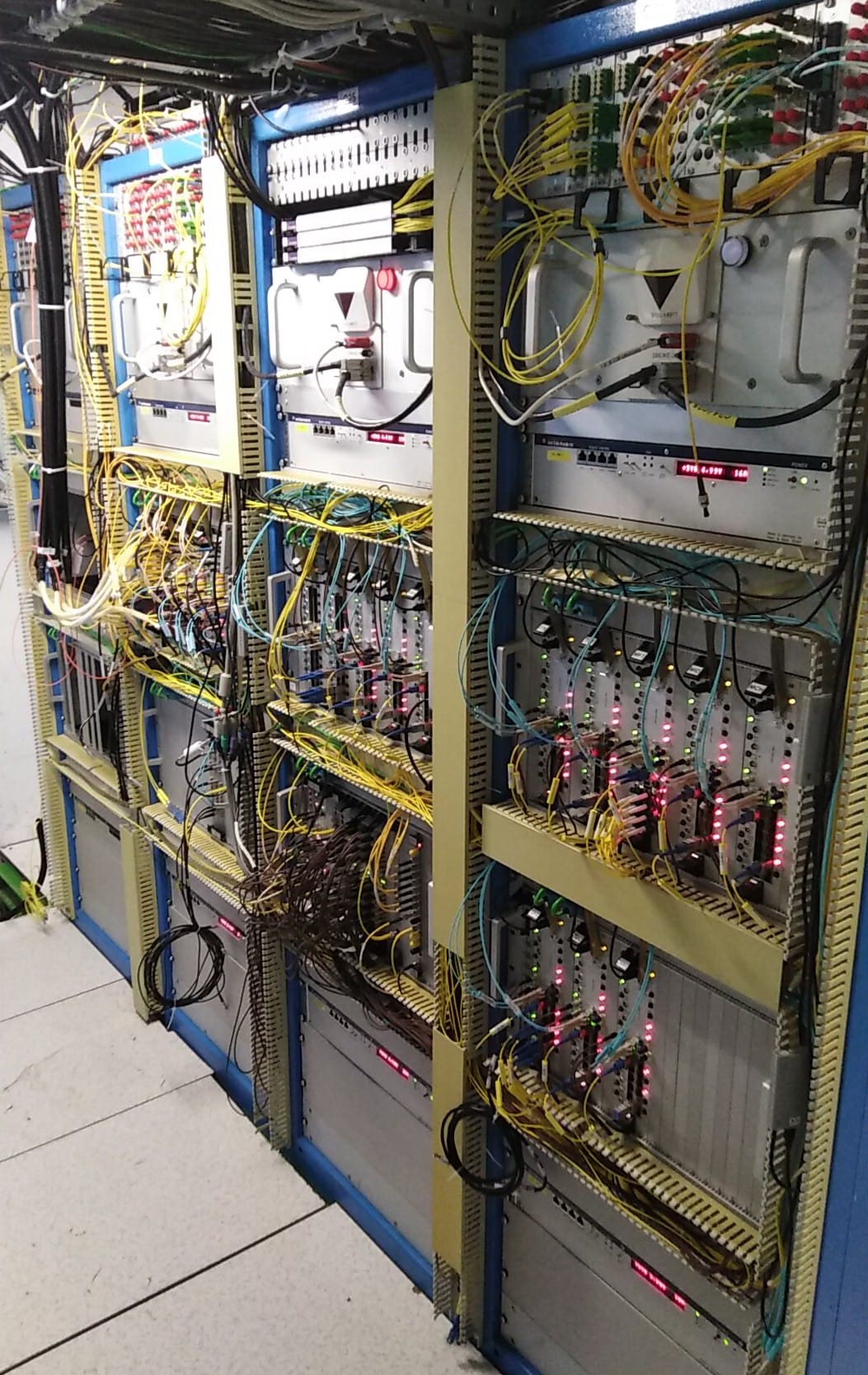}
	\end{minipage}	
	\caption{Central Trigger System overview with layout (left) and installation at point 2 (right).}
	\label{P2over}       
\end{figure}

The system was installed at LHC Point 2 at the end of September 2020. The CTP racks are as close as possible to the ALICE detectors and are located under the dipole magnet. The layout of the VME crates can be seen in Figure \ref{P2over} (left). C25B contains interfaces with the LHC mainly to receive LHC clock and orbit signals. The CTP and the FIT detector LTU are located in C25T (since the FIT provides a trigger input at LM latency it is as close as possible to CTP). Detectors using RD12 TTC protocol are located in C25T, where each LTU has a TTCex placed next to it. The CRU detectors are located in C24T and C24B. To monitor data on RD12 TTC optical links, an optical splitter is placed on top of C26 with a TTC interface test board in C26B. In case the ethernet connection cannot be used to reboot and update firmware and a hard reload using JTAG is required, a Network USB Hub - AnywhereUSB, placed on top of C24 and C26, is connected to each board through microUSB-JTAG. The final installation can be seen in Figure \ref{P2over} (right).

\section{Conclusions}
This paper describes the time and trigger distribution in ALICE for LHC Run 3. The design and implementation of the new ALICE CTS and connections and protocols are also discussed.
At the time of writing, all CTP and LTU boards have been produced and tested. The LTUs have been distributed to the ALICE detector groups and a full system has been installed at LHC Point 2. The ATB LTU firmware is nearly finished for  downstream data generation in standalone mode or for
passing the data received from the CTP. Development for the upstream data processing and passing it from detectors to the CTP has yet to be implemented. A first version of the CTP firmware, allowing global data taking in continuous mode is available. Development of the global continuous mode is the top priority. Software development is always subsequent to firmware implementation. However, further development work is still required for the ECS interface and continuous monitoring of the CTS.

\bibliography{bibliography}

\begin{thebibliography}{9}

\bibitem{alice}
K.~Aamodt, et~al. (ALICE Collaboration), Journal of Instrumentation \textbf{3},
  S08002 (2008)

\bibitem{LHC}
O.S. Brüning, P.~Collier, P.~Lebrun, S.~Myers, R.~Ostojic, J.~Poole,
  P.~Proudlock, CERN Yellow Reports: Monographs (CERN, Geneva, 2004),
  \urlstyle{tt}\url{https://cds.cern.ch/record/782076}

\bibitem{CTPTDR}
C.W. Fabjan, L.~Jirdén, V.~Lindestruth, L.~Riccati, D.~Rorich, P.~Van~de
  Vyvre, O.~Villalobos~Baillie, H.~de~Groot (ALICE Collaboration), Technical
  design report. ALICE (CERN, Geneva, 2004),
  \urlstyle{tt}\url{https://cds.cern.ch/record/684651}

\bibitem{RD12}
M.~Ashton, et~al. (RD12 Collaboration), Tech. rep., CERN, Geneva (2000),
  \urlstyle{tt}\url{https://cds.cern.ch/record/421208}

\bibitem{CRU}
J.~Mitra, S.~Khan, S.~Mukherjee, R.~Paul, \emph{{Common Readout Unit (CRU) - A
  new readout architecture for the ALICE experiment}}, Vol.~11 (2016)

\bibitem{PON}
E.B.S. Mendes, S.~Baron, D.M. Kolotouros, C.~Soos, F.~Vasey, \emph{{The 10G
  TTC-PON: challenges, solutions and performance}}, Vol.~12 (2017),
  \urlstyle{tt}\url{http://cds.cern.ch/record/2275139}

\bibitem{GBT}
P.~Moreira, R.~Ballabriga, S.~Baron, S.~Bonacini, O.~Cobanoglu, F.~Faccio,
  T.~Fedorov, , et~al., \emph{{The GBT Project}} (2009),
  \urlstyle{tt}\url{https://cds.cern.ch/record/1235836}

\bibitem{IPbus}
T.S. Williams (CMS Collaboration), \emph{{IPbus A flexible Ethernet-based
  control system for xTCA hardware}} (Geneva, 2014),
  \urlstyle{tt}\url{https://cds.cern.ch/record/2020872}

\bibitem{WinCC}
P.~Chochula, A.~Augustinus, P.~Bond, A.~Kurepin, M.~Lechman, O.~Pinazza,
  \emph{{The Evolution of the ALICE Detector Control System}} (2015),
  \urlstyle{tt}\url{https://cds.cern.ch/record/2213517}

\end{thebibliography}
%
%

\end{document}